\documentclass[aps,prl,twocolumn,floatfix,superscriptaddress,showpacs,amsfonts
,amssymb,amsmath,preprintnumbers]{revtex4-1}
\usepackage{bm}
\usepackage{graphicx}
\usepackage{color}
\usepackage{subfigure}
\usepackage[normalem]{ulem}
\newcommand{\be}{\begin{equation}}
\newcommand{\ee}{\end{equation}}
\newcommand{\beq}{\begin{equation}}
\newcommand{\eeq}{\end{equation}}

\newcommand\bea{\begin{eqnarray}}
\newcommand\eea{\end{eqnarray}}

\usepackage{flexisym}
\usepackage{wasysym}
\usepackage{natbib}
\usepackage{enumerate}

\usepackage{color}
\definecolor{darkblue}{rgb}{0.,0.,0.4}
\definecolor{darkred}{rgb}{0.5,0.,0.}
\usepackage[pdftex, colorlinks=true, linkcolor=darkblue,citecolor=blue,urlcolor=darkred]{hyperref}

\begin{document}
\title{Magnetic island merger as a mechanism for inverse magnetic energy transfer}

\author{Muni Zhou}
\affiliation{Plasma Science and Fusion Center$,$ Massachusetts Institute of Technology$,$ Cambridge$,$ MA 02139$,$ USA}
\author{Pallavi Bhat}
\affiliation{Plasma Science and Fusion Center$,$ Massachusetts Institute of Technology$,$ Cambridge$,$ MA 02139$,$ USA}
\author{Nuno F. Loureiro}
\affiliation{Plasma Science and Fusion Center$,$ Massachusetts Institute of Technology$,$ Cambridge$,$ MA 02139$,$ USA}
\author{Dmitri A. Uzdensky}
\affiliation{Center for Integrated Plasma Studies$,$ Physics Department$,$ UCB-390$,$ University of Colorado$,$ Boulder$,$ CO 80309$,$ USA}

\date{\today}

\begin{abstract}
Magnetic energy transfer from small to large scales due to successive magnetic island coalescence is investigated. 
A solvable analytical model is introduced and shown to correctly capture the evolution of the main quantities of interest, as borne out by numerical simulations.  
Magnetic reconnection is identified as the key mechanism enabling the inverse transfer, and setting its properties:
 magnetic energy decays as $\tilde t^{-1}$, where $\tilde t$ is time normalized to the (appropriately defined) reconnection timescale; and the correlation length of the field grows as $\tilde t^{1/2}$. 
The magnetic energy spectrum is self-similar, and evolves as $\propto \tilde t ^{-3/2}k^{-2}$, where the $k$-dependence is imparted by the formation of thin current sheets. 
\end{abstract}

\maketitle

\paragraph{Introduction.}
The transfer of magnetic energy from small to large spatial scales is a poorly understood plasma process of fundamental relevance to a variety of space and astrophysical environments.
It may, for example, play a critical role in the origin of large-scale galactic magnetic fields \cite{kulsrud2008origin}, by enabling kinetic-scale seed fields (e.g., Weibel~\cite{weibel1959} generated) to develop spatial coherence on larger, perhaps fluid, scales \cite{gruzinov2001gamma}. 
Ultimately, the questions are not only whether such an inverse cascade is possible, but also how rapid and efficient it is --- i.e., can an inverse cascade deliver significant amounts of  magnetic energy to scales where ambient turbulence may efficiently amplify it via turbulent dynamo processes?

Similarly motivated issues arise in the context of gamma-ray bursts (GRBs) where one wonders if Weibel-produced fields in the relativistic shock \cite{medvedev1999generation} can survive long enough to explain the observed powerful synchrotron emission~\cite{gruzinov2001gamma}. 
In the space-physics context, a frequently encountered question concerns the dynamic evolution of a complex, volume-filling ``sea'' of flux ropes or magnetic islands, e.g., in the solar wind and the outer heliosphere~\cite{manchester2004three,klimchuk2008highly,khabarova2015small,drake2012power}.

Past theoretical work on inverse magnetic energy transfer has mainly developed along two directions: the study of decaying turbulence, where the time evolution of a random, small-scale, initial field configuration is investigated~\cite[e.g.][]{biskamp1994dynamics,olesen1997inverse,biskamp2001two,zrake2014inverse,brandenburg2015nonhelical,zrake2016freely}, and the long-term evolution of Weibel-generated current filaments via their coalescence~\cite{medvedev2004long,kato2005saturation,katz2007self,fermo2010statistical,lyutikov_sironi_komissarov_porth_2017,gruzinov2001gamma}.
In this Letter, we build on concepts from both of these camps to present a conceptually new picture of inverse energy transfer which essentially relies on magnetic reconnection as the enabler of such a process. 

\paragraph{Hierarchical coalescence of magnetic islands.}
An analytically tractable model for inverse magnetic energy transfer is provided by a two-dimensional ensemble of identical magnetic islands whose evolution proceeds via their coalescence~\cite{finn1977coalescence}. 
Throughout the paper, for simplicity, we will adopt the resistive magnetohydrodynamic (MHD) framework, but we note that our ideas should qualitatively carry over to more advanced plasma descriptions.
We first assume that the (hierarchical) merging process occurs in discrete stages; at each stage (or generation, denoted by index~$n$), all islands are assumed circular and equal to each other. 

At any given $n$-th generation, a magnetic island can be characterized by its radius $R_n$ and the total flux it encloses, ~$\psi_n$. 
The typical magnetic field in the island, $B_n = \psi_n/R_n$, and the magnetic energy it contains, $\epsilon_n \simeq \pi R_n^2  B_n^2/(8\pi) =B_n^2/8\, R_n^2 = \psi_n^2/8$, can thus be determined. Other quantities of interest are the Alfv\'en velocity $v_{A,n}=B_n/\sqrt{4 \pi \rho}$ (the flow is assumed to be incompressible, thus the density $\rho$ is a constant); the number of islands per unit area $N_n$; and $\mathcal{E}_n = \epsilon_n N_n$, the total magnetic energy density of the system.

Island merger changes the above quantities. 
We will make two basic assumptions to determine the transition from one generation to the next~\cite{fermo2010statistical}. 
Firstly, the coalescence of two identical islands should conserve mass (and hence the area, due to the incompressibility assumption): two circular islands of radius~$R_n$ result in an island of radius $R_{n+1} = \sqrt{2} R_n$. 
Secondly, the magnetic flux should remain constant: $\psi_{n+1} = \psi_n$. 
The number density of islands, $N$, decreases by a factor of 2 through each stage; the evolution of other quantities can be determined straightforwardly from above conservation rules; 
e.g., $B_n$ and $v_{A,n}$ both decrease by~$\sqrt{2}$ (see also \cite{lyutikov_sironi_komissarov_porth_2017}).

To transition from this discrete description to a continuous time evolution, the life time for each island generation needs to be computed.
We consider coalescence to be a two-stage process: an initial island approach,  resulting from Lorentz attraction, proceeding at roughly the Alfv\'enic rate; and the subsequent reconnection of the two islands, taken to be much slower and thus dominating the overall merger duration.
We therefore express the merger time for $n$-th generation islands as $\tau_n \simeq \beta_{{\rm rec},n}^{-1}R_n/v_{A,n}$, 
where
$\beta_{{\rm rec},n}\ll 1$ is the dimensionless reconnection rate.

The main parameter controlling the reconnection regime, and hence $\beta_{{\rm rec},n}$, in resistive MHD is the Lundquist number set by the parameters of the merging islands: $S_n \equiv R_n v_{A,n}/\eta$, where $\eta$ is the (constant) magnetic diffusivity.  
In particular, if $S_n \lesssim 10^4$, reconnection proceeds in the Sweet-Parker (SP) regime~\cite{sweet_neutral_1958,parker_sweet_1957} with $\beta_{{\rm rec},n} \simeq S_n^{-1/2}$; if, instead, $S_n \gtrsim 10^4$, then reconnection proceeds in the plasmoid-dominated regime~\cite{loureiro2007instability,lapenta2008self,samtaney2009formation,bhattacharjee2009fast,huang2010scaling,uzdensky2010fast,loureiro2012magnetic,loureiro2013fast,loureiro_magnetic_2016} with $\beta_{{\rm rec},n}\simeq 0.01$.
Importantly, since $S_n=R_n v_{A,n} / \eta \propto R_n B_n \propto \psi_n$, which is 
preserved during mergers, we see that $S_n$, and thus $\beta_{{\rm rec},n}$, remain unchanged throughout the evolution ($S_n = S_0$).
This non-trivial result implies that the reconnection regime (SP or plasmoid-dominated) that governs the island mergers is set by the initial conditions\footnote{In the collisionless case $\beta_{\rm rec}\simeq 0.1$~\cite{cassak2017review} should also remain constant in time, even though the reconnection regime transitions from laminar to plasmoid-mediated as $R_n$ grows while the ion skin depth $d_i$ remains constant~\cite{ji2011phase} --- though the efficiency of coalescence may be decreased~\cite{karimabadi2011flux,stanier2015role}.}.

From the above recursive relations for $R_n$, $\psi_n$, and $\beta_{{\rm rec,}n}$, we find that the quantities evolve through generations as geometric progressions, resulting in:
\begin{equation}
\begin{aligned}
&\psi_n = \psi_0; \quad R_n = 2^{n/ 2}R_0; \quad B_n = 2^{-n/2}B_0;\\
&\mathcal{E}_n=2^{-n}\mathcal{E}_0; \quad N_n = 2^{-n} N_0;\quad \tau_n=2^n\tau_0.
\end{aligned}
\end{equation}

The time taken to reach the $n$th generation is
\begin{align}
&t_{n} = \sum_{k=0}^{n-1} \tau_k = \tau_0 \sum_{k=0}^{n-1} 2^k \approx \tau_0 2^{n}, \quad n\gg 1. \label{eq:time}
\end{align} 
Thus, the relationship between time and island generation $n$ is $t_n = \tau_0 \tilde{t} = \tau_0 2^n$, where $\tilde{t} \equiv t_n/\tau_0$. %(we note that the time normalization $\tau_0=\beta_{{\rm rec},0}^{-1}R_0/v_{A,0}$ can be very different for the SP and plasmoid reconnection regimes). 
This allows us to eliminate the index $n$ and obtain the explicit, continuous time dependence of the quantities of interest: 
\begin{align}
% \label{exponents}
 k &= k_0 \tilde{t}^{-1/2},  \quad B = B_0 \tilde{t}^{-1/2},\label{eq:expkB}\\
\mathcal{E} &= \mathcal{E}_0 \tilde{t}^{-1}, \quad N = N_0 \tilde{t}^{-1}, \quad \psi = \psi_0,\label{eq:expENpsi}
\end{align}
where $k \equiv 2\pi/R$. 

An alternative derivation of the scaling $B\sim t^{-1/2}$, Eq.~\eqref{eq:expkB}, can be obtained by expressing the time evolution of magnetic energy as $dB^2/dt\sim B^2/\tau_{\rm rec}$, where $\tau_{\rm rec}=\beta_{\rm rec}^{-1}R/v_{A}$ is the reconnection time.
The constancy both of the magnetic flux, $\psi = BR$, and of the reconnection rate, $\beta_{\rm rec}$, then implies that $\tau_{\rm rec}\propto B^{-2}$ and, therefore, $B\sim t^{-1/2}$.
Interestingly, the same scaling is obtained if we replace $\tau_{\rm rec}$ with $\tau_A=R/v_A$ as the characteristic timescale for magnetic energy evolution~\cite{biskamp1989dynamics,biskamp2001two,gruzinov2001gamma}.
Note, however, that this happens only because of the constancy of $\beta_{\rm rec}$ that we have derived: physically, the mechanism that dissipates magnetic energy is reconnection, and that is thus what sets its timescale.

The growing lengthscale and decreasing field strength, Eq.~\eqref{eq:expkB}, can also be interpreted from the perspective of dynamical renormalization. 
For an arbitrary scaling factor $l$, Eq.~\eqref{eq:expkB} is equivalent to the transformation:
\begin{equation}
\label{eq:scale_transformation}
    k \rightarrow l^{-1}k; \quad \tilde{t} \rightarrow l^{2}\tilde{t};\quad B \rightarrow l^{-1}B.
\end{equation}
It is reassuring --- and a confirmation of the consistency of our dynamical model --- that these relations are consistent with the general self-similar properties of the (unforced) MHD equations
\cite{olesen1997inverse,olesen2015inverse}; what we have shown, however, is that a physical process exists that enables such a rescaling.

\paragraph{Magnetic spectrum.}
The evolution of the system which we have just described is not, in fact, characterized by a single scale ($k_{\rm isl}$): the current sheets (of transverse scale $k_{\rm CS}$) which form during coalescence result in a wide magnetic energy Fourier spectrum, $k_{\rm isl}<k<k_{\rm CS}$, where, for SP reconnection, $k_{\rm isl}/k_{\rm CS} = S_0^{-1/2}$ (for $S_0<10^4$). 
Islands and sheets evolve together [$k_{\rm CS}(t) \propto k_{\rm isl}(t)$ since $S_0={\rm const}$], so, importantly, this entire scale range evolves on the same timescale.
Therefore, the magnetic power spectrum 
$U(k,\tilde{t})$ in this scale range,
\begin{equation}
    U(k,\tilde{t}) \equiv \frac{1}{8 \pi}\frac{2 \pi k}{(2 \pi)^2} \int d^2r
    e^{i \mathbf{k} \cdot\mathbf{r}} \langle \mathbf{B}(\mathbf{x},\tilde{t}) \cdot \mathbf{B}(\mathbf{x}+\mathbf{r},\tilde{t}) \rangle ,  \label{eq:defineU}
\end{equation}
transforms as $U(k/l,l^2\tilde{t}) = l^{-1}U(k,\tilde{t})$,
according to Eq.~\eqref{eq:scale_transformation}.
The spectra at different times are thus related by the scaling factor $l$, with a self-similar solution \cite{olesen1997inverse,olesen2015inverse}:
\begin{equation}
\label{eq:U_scalingfunction}
    U(k,\tilde{t})  =\tilde{t}^{-1/2}\bar{U}(k\tilde{t}^{1/2}),
\end{equation}
where $\bar{U}$ is a scaling function of the variable $k\tilde{t}^{1/2}$. 

In the particular case of a power-law spectrum,  $\Bar{U}(k\tilde{t}^{1/2}) \propto (k\tilde{t}^{1/2})^{-\gamma}$, the solution is:
\begin{equation}
\label{eq:U_k_t}
    	U(k,\tilde{t})\propto \tilde{t}^{-\alpha} k^{-\gamma},
\end{equation} 
where $2\alpha=\gamma+1$. 
In our system, the sharp magnetic field reversals at the current sheets are expected to lead to $\gamma=2$~\cite{burgers1948mathematical} (i.e., a $k^{-2}$ spectrum in the range $k_{\rm isl}<k<k_{\rm CS}$); and thus $\alpha=3/2$. The decay of energy density at any fixed wavenumber should then scale as $U_k(\tilde{t}) \propto \tilde{t}^{-3/2}$.

\begin{figure*}
\includegraphics[width=\textwidth,clip]{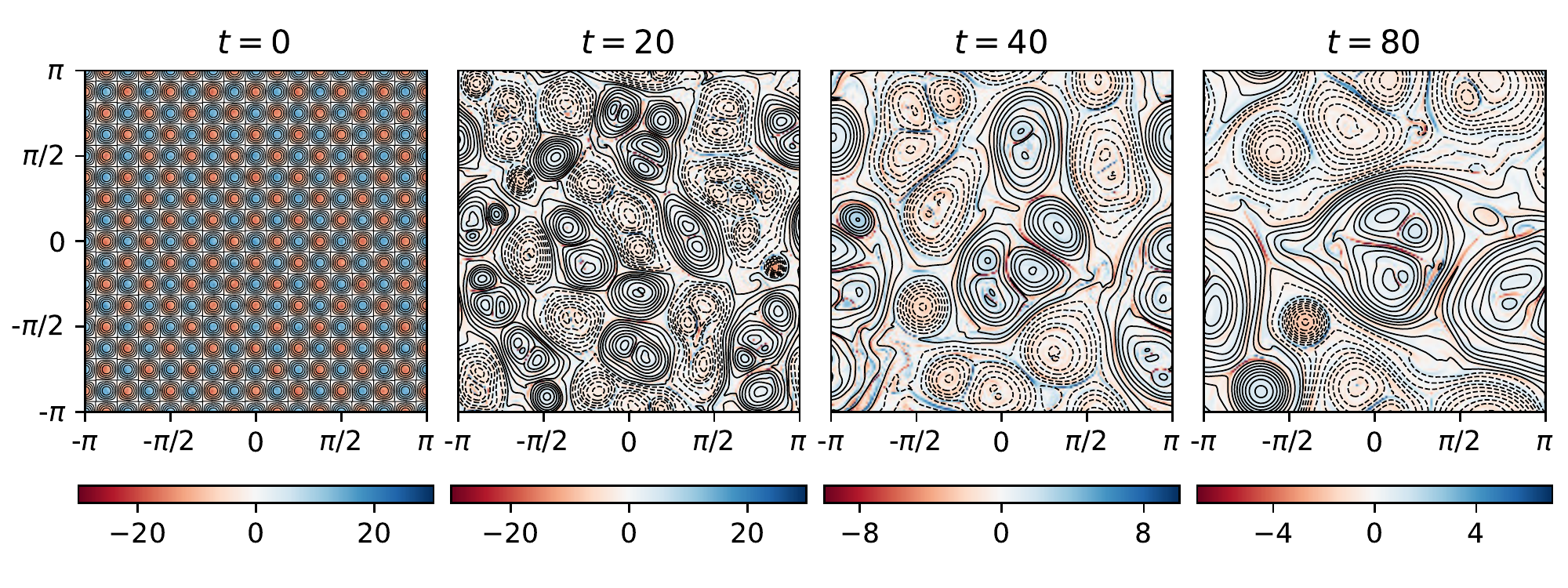}
\vspace*{-10mm}
\caption{Current density (colors) and magnetic flux (contours) at various times for the run with $S_0=1786$. }
\label{fig:contours}
\end{figure*}

\paragraph{Numerical Study.} 
To test the above results, we numerically solve the two-dimensional incompressible Reduced-MHD equations~\cite{kadomtsev1973nonlinear,strauss1976nonlinear,schekochihin2009astrophysical} using the pseudo-spectral code {\tt Viriato}~\cite{loureiro2016viriato}. 
In what follows, quantities are given in dimensionless form.
The domain is a periodic square box with sides of length $L=2\pi$. 
The initial equilibrium is described by the stream function $\phi(x,y)=0$ and the magnetic flux function $\psi(x,y) = \psi_0 \cos( k_0 x) \cos(k_0 y)$, yielding a $2k_0\times 2k_0$ static array of magnetic islands with opposite polarities (Fig.~\ref{fig:contours}, left panel). 
In all runs we choose $k_0=8$, and thus $R_0=L/4k_0=\pi/16$.
We further set $\psi_0 k_0=1$, implying $B_0\equiv\psi_0/R_0=2/\pi$. 
This initial equilibrium is perturbed by small-amplitude, spatially random noise 
to initiate the evolution.
We perform a series of runs for different values of resistivity $\eta \in \{1\times 10^{-3}, 7 \times 10^{-4},5 \times 10^{-4}, 3 \times 10^{-4}, 1\times 10^{-4}, 7\times 10^{-5}\}$, which correspond to the initial island-scale Lundquist number $S_0\equiv R_0 v_{A,0}/\eta \in \{125, 179, 250,417,1250,1786\}$. 
Viscosity is set equal to resistivity. 
We use $8192^2$ grid points for $S_0=1786$; $4096^2$ for $S_0=1250,417$; and $2048^2$ for $S_0=250,179,125$. 
The widths of initial (SP) current sheets are resolved with 3 or 4 grid points in all cases.
Since $S_0<10^4$ in all runs, reconnection should proceed in the SP regime and no (secondary) plasmoids are expected to arise; visual inspection of our simulations confirms this.
Fig.~\ref{fig:contours} shows the configuration of the system with $S_0=1786$ at different times. 
As expected, island mergers lead to the progressive formation of ever larger structures.

\begin{figure}[htp]
    \includegraphics[width=0.45\textwidth]{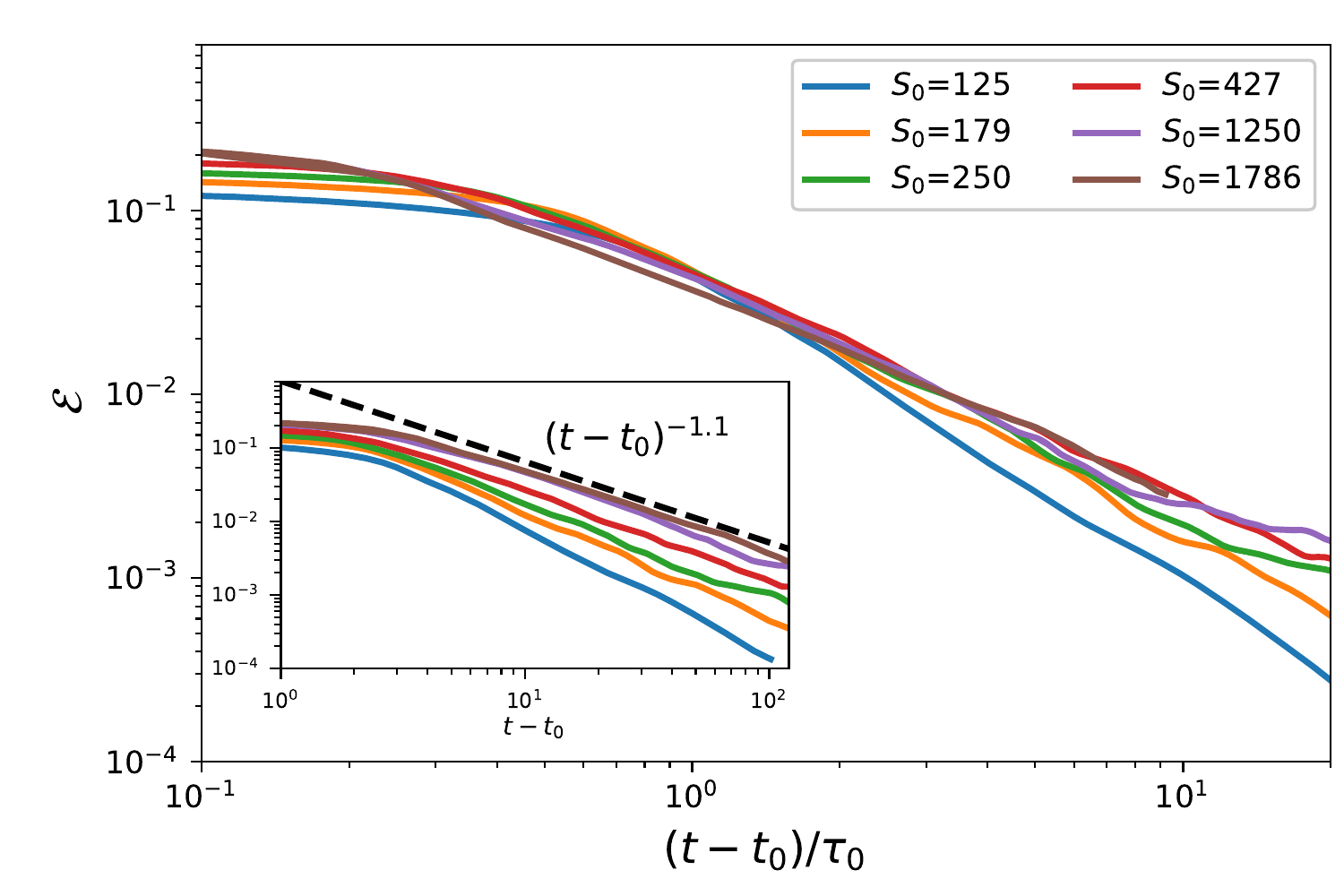}
    \vspace*{-4mm}
    \caption{Time evolution of total magnetic energy ($\mathcal{E}$) for all values of~$S_0$. Time axis in the main figure is normalized to the reconnection timescale  $\tau_0=S_0^{1/2}(\pi^2/32)$ (in code units).  The inset figure shows same data {\it vs.} time in code units. The power-law fit for $S_0=1786$ is indicated for reference. }
    \label{fig:UtNt}
\end{figure}
\begin{figure}[htp]
    \centering
    \includegraphics[width=3.375 in]{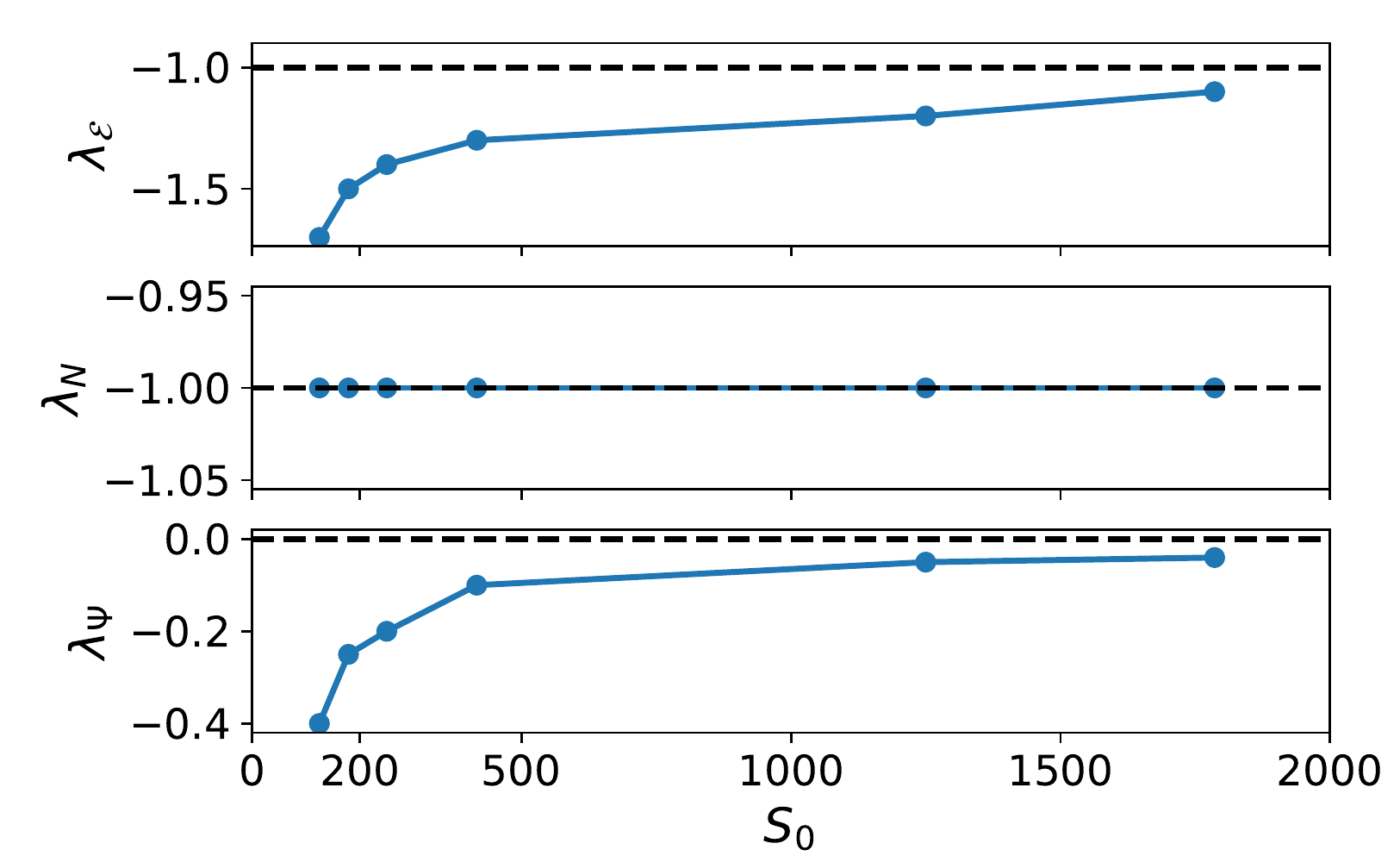}
    \vspace*{-7mm}
    \caption{Power-law exponents as functions of initial Lundquist number $S_0$. Dashed lines represent the predictions of our model, Eq. \eqref{eq:expENpsi}.}
    \label{fig:Es}
\end{figure}

In Fig.~\ref{fig:UtNt} we plot the time evolution of the total magnetic energy $\mathcal{E}$ for all values of~$S_0$.
After an initial transient period (represented by a time offset $t_0$; for $S_0=1786$ it is $t_0=4$) 
the system enters a prolonged stage of self-similar evolution with power-law-in-time behavior; other quantities, such as the number of islands\footnote{The number of islands is numerically determined by diagnosing the O-points and X-points of the system, identified with the maximum/minimum and saddle points of $\psi(x,y)$~\cite[e.g.,][]{servidio2009magnetic}.}, 
$N(t)$, or the spatial maximum of the flux, $\psi_{\rm max}(t)$, behave similarly. 
We fit this data to functions of the form $(t-t_0)^\lambda$.
The measured power-law indices, $\lambda_{\mathcal{E}},~\lambda_N$, and $\lambda_\psi$ are found to converge to the predictions of our hierarchical model, Eq.~\eqref{eq:expENpsi}, as $S_0$ increases, as shown in Fig.~\ref{fig:Es}.
This suggests that the hierarchical model can indeed capture the basic dynamics of the merging system.
We also find that the time offset $t_0$ increases with~$S_0$, consistent with the expected scaling of the reconnection rate. 

Additionally, Fig.~\ref{fig:UtNt} clearly demonstrates that the characteristic timescale for the magnetic energy evolution is the reconnection time, $\tau_0$. 
This is evidenced by the approximate collapse of all curves in the main plot, where the time axis is normalized to $\tau_0$, but not in the inset figure, where time is in code units.

\begin{figure}[htp]
    \includegraphics[clip,width=0.45\textwidth]{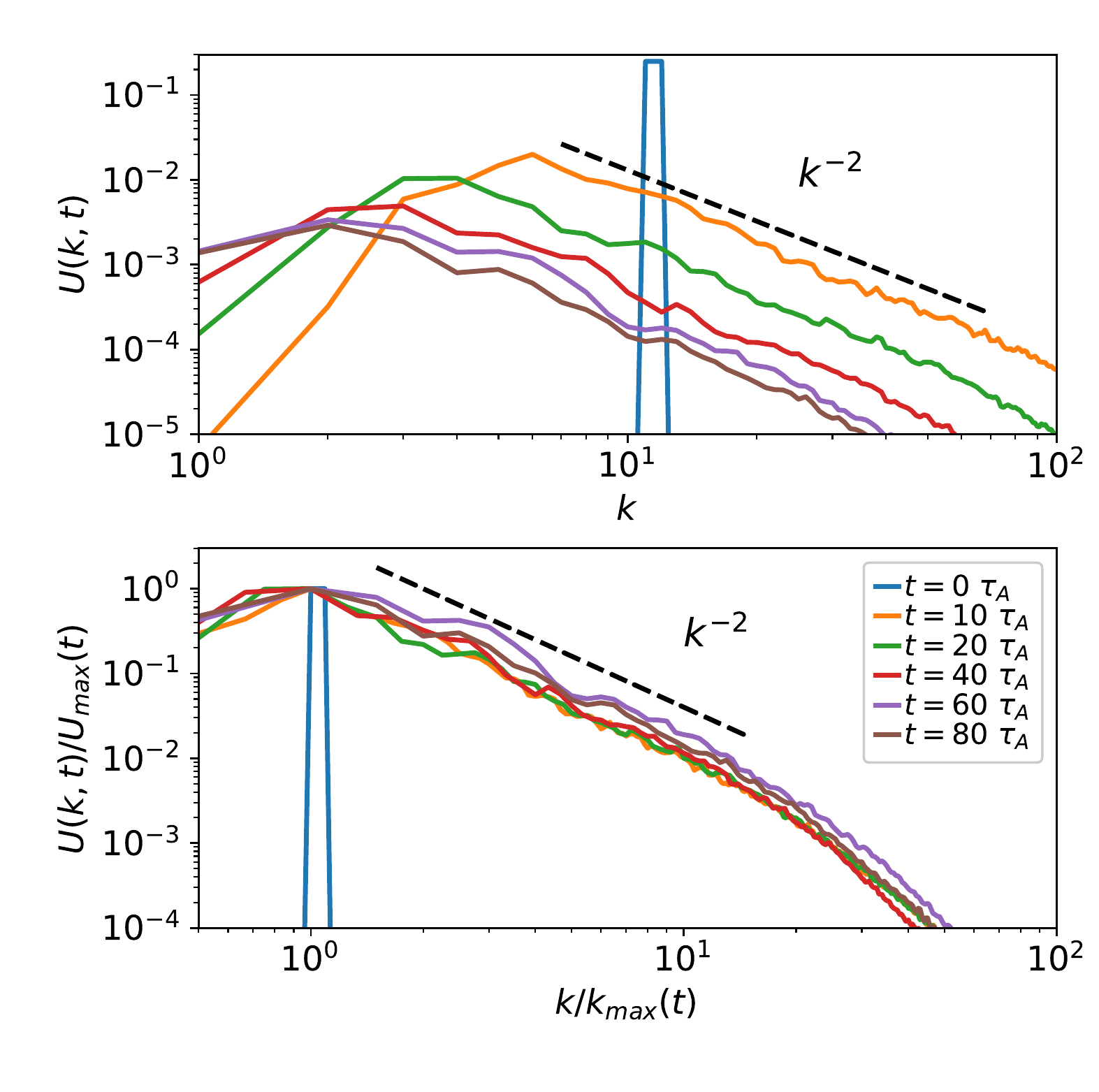}
    \vspace*{-6mm}
    \caption{Raw (top) and normalized (bottom) magnetic power spectra, for $S_0=1786$. A $k^{-2}$ slope is shown for reference.}
    \label{fig:Spectra}
\end{figure}

Fig.~\ref{fig:Spectra} (top panel) shows the magnetic spectrum $U(k,t)$ at different moments of time for the $S_0=1786$ run. 
As is visually intuited from Fig.~\ref{fig:contours}, we observe that the peak of the spectrum moves to larger scales, while retaining an overall similar shape. 
To the right of the peak, these spectra exhibit power-law behavior (an inertial range), with a slope that is well approximated by the index $\gamma=2$ (in agreement with ~\cite{brandenburg2015nonhelical,zrake2014inverse}).
We think that this index is due to the presence of thin current sheets~\cite{burgers1948mathematical}; indeed, a $k^{-2}$ slope forms even before any coalescence has taken place, and thus it cannot be yielded by the magnetic island distribution.
The kinetic energy spectrum (not shown) exhibits a peak at roughly the same wavenumber as the magnetic energy, but follows a shallower power-law, $\sim k^{-1}$. 
This is consistent with the notion that kinetic energy in the current sheets is dominated by the (Alfv\'enic) outflows (whose spatial profile~\cite{loureiro2013plasmoid} yields a flat spectrum), plus background flows (both inside and outside the magnetic islands) on the scale of the dominant islands\footnote{We also observe in our simulations that kinetic energy decays as $u^2\sim t^{-1}$, consistent with $u\sim B$.}.

The self-similarity of the magnetic spectra is clearly demonstrated in the bottom panel of Fig.~\ref{fig:Spectra}, where we normalize the spectra to their respective maximum values at each moment of time, $U_{\rm max}(t)$, and the wavenumbers to the values $k_{\rm max}(t)$ at which $U_{\rm max}(t)$ are attained. 
As seen, all curves essentially collapse onto the same distribution, implying that $U(k,t)$ can be factorized as $ U(k,t) = U_{\rm max}(t) \bar{U}(k/k_{\rm max})$.

\begin{figure}[htp]
    \includegraphics[width=0.45\textwidth]{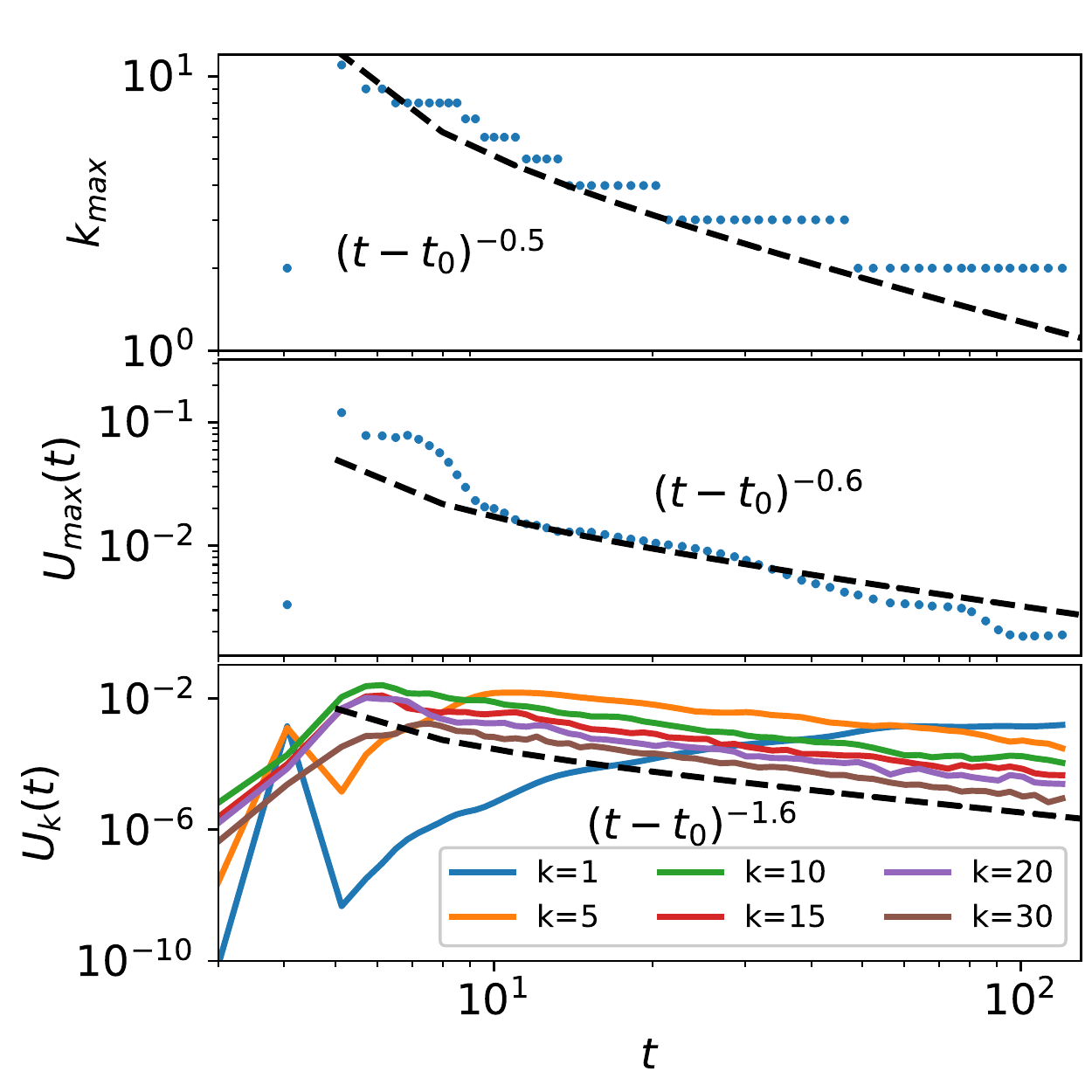}
    \vspace*{-4mm}
    \caption{Time evolution of $k_{\rm max}$ (top), $U_{\rm max}$ (middle) and $U_k$ (for selected values of $k$; bottom), for $S_0=1786$.}
    \label{fig:kmaxUmax}
\end{figure}

We also observe that $k_{\rm max}$ and $U_{\rm max}$ are roughly power-law functions of time, $k_{\rm max} \propto \Delta t^{-\beta}$ and $U_{\rm max}\propto \Delta t^{-\theta}$, where $\Delta t \equiv t-t_0$, as shown in the top two panels of Fig.~\ref{fig:kmaxUmax}. Hence $U(k,t)$ can be expressed as $U(k,t) \propto \Delta t^{-\theta} \bar{U}(k \Delta t^{\beta})$, where $\bar{U}(k/k_{\rm max})$ becomes a universal scaling function of the variable~$k\Delta t^\beta$, consistent with Eq.~\eqref{eq:U_scalingfunction}.
As noted above, the spectra exhibit an inertial range [$k>k_{\rm max}(t)$] with a power-law dependence on $k$: $U\propto k^{-\gamma}$ (Fig.~\ref{fig:Spectra}, top panel).
Therefore, in the inertial range, we have the power-law scaling function: $\bar{U}(k\Delta t^{\beta})\propto (k\Delta t^{\beta})^{-\gamma}$, and the power-law time dependence of magnetic spectral energy density at any given~$k$: $U_k(t) \propto \Delta t^{-\alpha}$ (Fig.~\ref{fig:kmaxUmax}, bottom panel), leading to the general expression for the spectrum: $U(k,t) \propto k^{-\gamma}\Delta t^{-\alpha}$, where $\alpha=\gamma\beta+\theta$ (see also~\cite{zrake2014inverse}).

The measured values of all the indices as well as the relation between $\alpha$ and $\gamma$, can be compared with our model [Eqs.~\eqref{eq:U_scalingfunction}--\eqref{eq:U_k_t}].
The results for different simulations are summarized in Table~\ref{table:exp}, where we also include a $4096^2$ simulation performed with hyper-dissipation --- this enables us to extend the range of dissipation-free scales as much as possible. 
We observe that as $S_0$ increases,  the exponents approach our theoretical predictions.

Lastly, we have performed one run ($\eta=10^{-4}, 4096^2$ grid cells) where the initial condition is instead a Gaussian-random magnetic field, with a spectrum narrowly peaked around~$k_0 = 8$. 
We observe that the power-law exponents obtained in this run are close to those in the run with same~$\eta$ but with the periodic-islands initial configuration. 
This indicates that our reconnection-based hierarchical model can describe the inverse magnetic energy transfer in a more general decaying 2D turbulent system.
\begin{table}[h!]
\begin{tabular}{ |p{1.6cm}|p{1cm} p{1cm} p{1cm} p{1cm}|  }
 \hline
 $S_0$ & $\beta$ & $\theta$& $\alpha$& $\gamma$  \\
 \hline \hline
 $125$ &0.5 &1.5&3.0&3.5\\
 $179$ &0.6 &1.3&2.1&2.9\\
 $250$ &0.6 &1.3&2.0&2.1\\
 $417$ &0.6 &1.2&1.9&2.1\\
 $1250$ &0.5 &1.1&1.8&2.0\\
 $1786$ &0.5 &0.6&1.6&2.0\\
 Hyper-diss &0.5&0.5&1.5&2.0\\
 \hline \hline
 Theory  &0.5 &0.5& 1.5 &2.0  \\
 \hline
\end{tabular}
\caption{Variation of exponents with initial Lundquist number $S_0$, compared with the prediction from the hierarchical model.}
\label{table:exp}
\end{table}%

\paragraph{Conclusion.}
We have introduced a solvable analytic model to describe the inverse transfer of magnetic energy arising from the hierarchical merger of magnetic islands via magnetic reconnection.
We have also carried out direct numerical simulations which show good agreement with the predictions of the model, thereby identifying reconnection as the mechanism that sets the properties (including, importantly, the timescale) of such inverse energy transfer. 
These results --- and, more generally, the notion of reconnection as the enabler of inverse energy transfer --- are of broad applicability to various space and astrophysical environments.
They may, for example, pave the way for understanding the long-term evolution of kinetic-scale seed magnetic fields: a longstanding problem in plasma astrophysics with direct implications to GRBs and galactic magnetogenesis.

\paragraph{Acknowledgements.}
This work was supported by NSF CAREER award No.~1654168 (MZ and NFL), NSF-DOE Partnership in Basic Plasma Science and Engineering Award No.~DE-SC0016215 (PB), and NSF grants AST-1411879 and AST-1806084 and NASA ATP grants NNX16AB28G and NNX17AK57G (DAU).  
D.A.U. gratefully acknowledges the hospitality of the Institute for Advanced Study and the support from the Ambrose Monell Foundation.
The authors thank James~A.~Klimchuk, Giovanni~Lapenta, Alexander~A.~Schekochihin and Liujun Zou for insightful discussions.
The simulations presented in this paper were performed on the MIT-PSFC partition of the Engaging cluster at the MGHPCC facility, funded by DOE award No.~DE-FG02-91-ER54109.

\bibliography{ref}

\begin{thebibliography}{47}%
\makeatletter
\providecommand \@ifxundefined [1]{%
 \@ifx{#1\undefined}
}%
\providecommand \@ifnum [1]{%
 \ifnum #1\expandafter \@firstoftwo
 \else \expandafter \@secondoftwo
 \fi
}%
\providecommand \@ifx [1]{%
 \ifx #1\expandafter \@firstoftwo
 \else \expandafter \@secondoftwo
 \fi
}%
\providecommand \natexlab [1]{#1}%
\providecommand \enquote  [1]{``#1''}%
\providecommand \bibnamefont  [1]{#1}%
\providecommand \bibfnamefont [1]{#1}%
\providecommand \citenamefont [1]{#1}%
\providecommand \href@noop [0]{\@secondoftwo}%
\providecommand \href [0]{\begingroup \@sanitize@url \@href}%
\providecommand \@href[1]{\@@startlink{#1}\@@href}%
\providecommand \@@href[1]{\endgroup#1\@@endlink}%
\providecommand \@sanitize@url [0]{\catcode `\\12\catcode `\$12\catcode
  `\&12\catcode `\#12\catcode `\^12\catcode `\_12\catcode `\%12\relax}%
\providecommand \@@startlink[1]{}%
\providecommand \@@endlink[0]{}%
\providecommand \url  [0]{\begingroup\@sanitize@url \@url }%
\providecommand \@url [1]{\endgroup\@href {#1}{\urlprefix }}%
\providecommand \urlprefix  [0]{URL }%
\providecommand \Eprint [0]{\href }%
\providecommand \doibase [0]{http://dx.doi.org/}%
\providecommand \selectlanguage [0]{\@gobble}%
\providecommand \bibinfo  [0]{\@secondoftwo}%
\providecommand \bibfield  [0]{\@secondoftwo}%
\providecommand \translation [1]{[#1]}%
\providecommand \BibitemOpen [0]{}%
\providecommand \bibitemStop [0]{}%
\providecommand \bibitemNoStop [0]{.\EOS\space}%
\providecommand \EOS [0]{\spacefactor3000\relax}%
\providecommand \BibitemShut  [1]{\csname bibitem#1\endcsname}%
\let\auto@bib@innerbib\@empty
%</preamble>
\bibitem [{\citenamefont {Kulsrud}\ and\ \citenamefont
  {Zweibel}(2008)}]{kulsrud2008origin}%
  \BibitemOpen
  \bibfield  {author} {\bibinfo {author} {\bibfnamefont {R.~M.}\ \bibnamefont
  {Kulsrud}}\ and\ \bibinfo {author} {\bibfnamefont {E.~G.}\ \bibnamefont
  {Zweibel}},\ }\href@noop {} {\bibfield  {journal} {\bibinfo  {journal}
  {Reports on Progress in Physics}\ }\textbf {\bibinfo {volume} {71}},\
  \bibinfo {pages} {046901} (\bibinfo {year} {2008})}\BibitemShut {NoStop}%
\bibitem [{\citenamefont {Weibel}(1959)}]{weibel1959}%
  \BibitemOpen
  \bibfield  {author} {\bibinfo {author} {\bibfnamefont {E.~S.}\ \bibnamefont
  {Weibel}},\ }\href@noop {} {\bibfield  {journal} {\bibinfo  {journal}
  {Physical Review Letters}\ }\textbf {\bibinfo {volume} {2}},\ \bibinfo
  {pages} {83} (\bibinfo {year} {1959})}\BibitemShut {NoStop}%
\bibitem [{\citenamefont {Gruzinov}(2001)}]{gruzinov2001gamma}%
  \BibitemOpen
  \bibfield  {author} {\bibinfo {author} {\bibfnamefont {A.}~\bibnamefont
  {Gruzinov}},\ }\href@noop {} {\bibfield  {journal} {\bibinfo  {journal} {The
  Astrophysical Journal Letters}\ }\textbf {\bibinfo {volume} {563}},\ \bibinfo
  {pages} {L15} (\bibinfo {year} {2001})}\BibitemShut {NoStop}%
\bibitem [{\citenamefont {Medvedev}\ and\ \citenamefont
  {Loeb}(1999)}]{medvedev1999generation}%
  \BibitemOpen
  \bibfield  {author} {\bibinfo {author} {\bibfnamefont {M.~V.}\ \bibnamefont
  {Medvedev}}\ and\ \bibinfo {author} {\bibfnamefont {A.}~\bibnamefont
  {Loeb}},\ }\href@noop {} {\bibfield  {journal} {\bibinfo  {journal} {The
  Astrophysical Journal}\ }\textbf {\bibinfo {volume} {526}},\ \bibinfo {pages}
  {697} (\bibinfo {year} {1999})}\BibitemShut {NoStop}%
\bibitem [{\citenamefont {Manchester~IV}\ \emph {et~al.}()\citenamefont
  {Manchester~IV}, \citenamefont {Gombosi}, \citenamefont {Roussev},
  \citenamefont {De~Zeeuw}, \citenamefont {Sokolov}, \citenamefont {Powell},
  \citenamefont {Tóth},\ and\ \citenamefont {Opher}}]{manchester2004three}%
  \BibitemOpen
  \bibfield  {author} {\bibinfo {author} {\bibfnamefont {W.~B.}\ \bibnamefont
  {Manchester~IV}}, \bibinfo {author} {\bibfnamefont {T.~I.}\ \bibnamefont
  {Gombosi}}, \bibinfo {author} {\bibfnamefont {I.}~\bibnamefont {Roussev}},
  \bibinfo {author} {\bibfnamefont {D.~L.}\ \bibnamefont {De~Zeeuw}}, \bibinfo
  {author} {\bibfnamefont {I.~V.}\ \bibnamefont {Sokolov}}, \bibinfo {author}
  {\bibfnamefont {K.~G.}\ \bibnamefont {Powell}}, \bibinfo {author}
  {\bibfnamefont {G.}~\bibnamefont {Tóth}}, \ and\ \bibinfo {author}
  {\bibfnamefont {M.}~\bibnamefont {Opher}},\ }\href@noop {} {\bibfield
  {journal} {\bibinfo  {journal} {Journal of Geophysical Research: Space
  Physics}\ }\textbf {\bibinfo {volume} {109}}}\BibitemShut {NoStop}%
\bibitem [{\citenamefont {{Klimchuk}}\ \emph {et~al.}(2008)\citenamefont
  {{Klimchuk}}, \citenamefont {{Patsourakos}},\ and\ \citenamefont
  {{Cargill}}}]{klimchuk2008highly}%
  \BibitemOpen
  \bibfield  {author} {\bibinfo {author} {\bibfnamefont {J.~A.}\ \bibnamefont
  {{Klimchuk}}}, \bibinfo {author} {\bibfnamefont {S.}~\bibnamefont
  {{Patsourakos}}}, \ and\ \bibinfo {author} {\bibfnamefont {P.~J.}\
  \bibnamefont {{Cargill}}},\ }\href@noop {} {\bibfield  {journal} {\bibinfo
  {journal} {\apj}\ }\textbf {\bibinfo {volume} {682}},\ \bibinfo {pages}
  {1351} (\bibinfo {year} {2008})}\BibitemShut {NoStop}%
\bibitem [{\citenamefont {Khabarova}\ \emph {et~al.}(2015)\citenamefont
  {Khabarova}, \citenamefont {Zank}, \citenamefont {Li}, \citenamefont
  {Le~Roux}, \citenamefont {Webb}, \citenamefont {Dosch},\ and\ \citenamefont
  {Malandraki}}]{khabarova2015small}%
  \BibitemOpen
  \bibfield  {author} {\bibinfo {author} {\bibfnamefont {O.}~\bibnamefont
  {Khabarova}}, \bibinfo {author} {\bibfnamefont {G.}~\bibnamefont {Zank}},
  \bibinfo {author} {\bibfnamefont {G.}~\bibnamefont {Li}}, \bibinfo {author}
  {\bibfnamefont {J.}~\bibnamefont {Le~Roux}}, \bibinfo {author} {\bibfnamefont
  {G.}~\bibnamefont {Webb}}, \bibinfo {author} {\bibfnamefont {A.}~\bibnamefont
  {Dosch}}, \ and\ \bibinfo {author} {\bibfnamefont {O.}~\bibnamefont
  {Malandraki}},\ }\href@noop {} {\bibfield  {journal} {\bibinfo  {journal}
  {The Astrophysical Journal}\ }\textbf {\bibinfo {volume} {808}},\ \bibinfo
  {pages} {181} (\bibinfo {year} {2015})}\BibitemShut {NoStop}%
\bibitem [{\citenamefont {{Drake}}\ \emph {et~al.}(2012)\citenamefont
  {{Drake}}, \citenamefont {{Swisdak}},\ and\ \citenamefont
  {{Fermo}}}]{drake2012power}%
  \BibitemOpen
  \bibfield  {author} {\bibinfo {author} {\bibfnamefont {J.~F.}\ \bibnamefont
  {{Drake}}}, \bibinfo {author} {\bibfnamefont {M.}~\bibnamefont {{Swisdak}}},
  \ and\ \bibinfo {author} {\bibfnamefont {R.}~\bibnamefont {{Fermo}}},\
  }\href@noop {} {\bibfield  {journal} {\bibinfo  {journal} {The Astrophysical
  Journal Letters}\ }\textbf {\bibinfo {volume} {763}},\ \bibinfo {pages} {L5}
  (\bibinfo {year} {2012})}\BibitemShut {NoStop}%
\bibitem [{\citenamefont {Biskamp}\ and\ \citenamefont
  {Bremer}(1994)}]{biskamp1994dynamics}%
  \BibitemOpen
  \bibfield  {author} {\bibinfo {author} {\bibfnamefont {D.}~\bibnamefont
  {Biskamp}}\ and\ \bibinfo {author} {\bibfnamefont {U.}~\bibnamefont
  {Bremer}},\ }\href@noop {} {\bibfield  {journal} {\bibinfo  {journal}
  {Physical Review Letters}\ }\textbf {\bibinfo {volume} {72}},\ \bibinfo
  {pages} {3819} (\bibinfo {year} {1994})}\BibitemShut {NoStop}%
\bibitem [{\citenamefont {Olesen}(1997)}]{olesen1997inverse}%
  \BibitemOpen
  \bibfield  {author} {\bibinfo {author} {\bibfnamefont {P.}~\bibnamefont
  {Olesen}},\ }\href@noop {} {\bibfield  {journal} {\bibinfo  {journal}
  {Physics Letters B}\ }\textbf {\bibinfo {volume} {398}},\ \bibinfo {pages}
  {321} (\bibinfo {year} {1997})}\BibitemShut {NoStop}%
\bibitem [{\citenamefont {Biskamp}\ and\ \citenamefont
  {Schwarz}(2001)}]{biskamp2001two}%
  \BibitemOpen
  \bibfield  {author} {\bibinfo {author} {\bibfnamefont {D.}~\bibnamefont
  {Biskamp}}\ and\ \bibinfo {author} {\bibfnamefont {E.}~\bibnamefont
  {Schwarz}},\ }\href@noop {} {\bibfield  {journal} {\bibinfo  {journal}
  {Physics of Plasmas}\ }\textbf {\bibinfo {volume} {8}},\ \bibinfo {pages}
  {3282} (\bibinfo {year} {2001})}\BibitemShut {NoStop}%
\bibitem [{\citenamefont {Zrake}(2014)}]{zrake2014inverse}%
  \BibitemOpen
  \bibfield  {author} {\bibinfo {author} {\bibfnamefont {J.}~\bibnamefont
  {Zrake}},\ }\href@noop {} {\bibfield  {journal} {\bibinfo  {journal} {The
  Astrophysical Journal Letters}\ }\textbf {\bibinfo {volume} {794}},\ \bibinfo
  {pages} {L26} (\bibinfo {year} {2014})}\BibitemShut {NoStop}%
\bibitem [{\citenamefont {Brandenburg}\ \emph {et~al.}(2015)\citenamefont
  {Brandenburg}, \citenamefont {Kahniashvili},\ and\ \citenamefont
  {Tevzadze}}]{brandenburg2015nonhelical}%
  \BibitemOpen
  \bibfield  {author} {\bibinfo {author} {\bibfnamefont {A.}~\bibnamefont
  {Brandenburg}}, \bibinfo {author} {\bibfnamefont {T.}~\bibnamefont
  {Kahniashvili}}, \ and\ \bibinfo {author} {\bibfnamefont {A.~G.}\
  \bibnamefont {Tevzadze}},\ }\href@noop {} {\bibfield  {journal} {\bibinfo
  {journal} {Physical Review Letters}\ }\textbf {\bibinfo {volume} {114}},\
  \bibinfo {pages} {075001} (\bibinfo {year} {2015})}\BibitemShut {NoStop}%
\bibitem [{\citenamefont {Zrake}\ and\ \citenamefont
  {East}(2016)}]{zrake2016freely}%
  \BibitemOpen
  \bibfield  {author} {\bibinfo {author} {\bibfnamefont {J.}~\bibnamefont
  {Zrake}}\ and\ \bibinfo {author} {\bibfnamefont {W.~E.}\ \bibnamefont
  {East}},\ }\href@noop {} {\bibfield  {journal} {\bibinfo  {journal} {The
  Astrophysical Journal}\ }\textbf {\bibinfo {volume} {817}},\ \bibinfo {pages}
  {89} (\bibinfo {year} {2016})}\BibitemShut {NoStop}%
\bibitem [{\citenamefont {Medvedev}\ \emph {et~al.}(2004)\citenamefont
  {Medvedev}, \citenamefont {Fiore}, \citenamefont {Fonseca}, \citenamefont
  {Silva},\ and\ \citenamefont {Mori}}]{medvedev2004long}%
  \BibitemOpen
  \bibfield  {author} {\bibinfo {author} {\bibfnamefont {M.~V.}\ \bibnamefont
  {Medvedev}}, \bibinfo {author} {\bibfnamefont {M.}~\bibnamefont {Fiore}},
  \bibinfo {author} {\bibfnamefont {R.~A.}\ \bibnamefont {Fonseca}}, \bibinfo
  {author} {\bibfnamefont {L.~O.}\ \bibnamefont {Silva}}, \ and\ \bibinfo
  {author} {\bibfnamefont {W.~B.}\ \bibnamefont {Mori}},\ }\href@noop {}
  {\bibfield  {journal} {\bibinfo  {journal} {The Astrophysical Journal
  Letters}\ }\textbf {\bibinfo {volume} {618}},\ \bibinfo {pages} {L75}
  (\bibinfo {year} {2004})}\BibitemShut {NoStop}%
\bibitem [{\citenamefont {Kato}(2005)}]{kato2005saturation}%
  \BibitemOpen
  \bibfield  {author} {\bibinfo {author} {\bibfnamefont {T.~N.}\ \bibnamefont
  {Kato}},\ }\href@noop {} {\bibfield  {journal} {\bibinfo  {journal} {Physics
  of plasmas}\ }\textbf {\bibinfo {volume} {12}},\ \bibinfo {pages} {080705}
  (\bibinfo {year} {2005})}\BibitemShut {NoStop}%
\bibitem [{\citenamefont {Katz}\ \emph {et~al.}(2007)\citenamefont {Katz},
  \citenamefont {Keshet},\ and\ \citenamefont {Waxman}}]{katz2007self}%
  \BibitemOpen
  \bibfield  {author} {\bibinfo {author} {\bibfnamefont {B.}~\bibnamefont
  {Katz}}, \bibinfo {author} {\bibfnamefont {U.}~\bibnamefont {Keshet}}, \ and\
  \bibinfo {author} {\bibfnamefont {E.}~\bibnamefont {Waxman}},\ }\href@noop {}
  {\bibfield  {journal} {\bibinfo  {journal} {The Astrophysical Journal}\
  }\textbf {\bibinfo {volume} {655}},\ \bibinfo {pages} {375} (\bibinfo {year}
  {2007})}\BibitemShut {NoStop}%
\bibitem [{\citenamefont {Fermo}\ \emph {et~al.}(2010)\citenamefont {Fermo},
  \citenamefont {Drake},\ and\ \citenamefont {Swisdak}}]{fermo2010statistical}%
  \BibitemOpen
  \bibfield  {author} {\bibinfo {author} {\bibfnamefont {R.}~\bibnamefont
  {Fermo}}, \bibinfo {author} {\bibfnamefont {J.}~\bibnamefont {Drake}}, \ and\
  \bibinfo {author} {\bibfnamefont {M.}~\bibnamefont {Swisdak}},\ }\href@noop
  {} {\bibfield  {journal} {\bibinfo  {journal} {Physics of Plasmas}\ }\textbf
  {\bibinfo {volume} {17}},\ \bibinfo {pages} {010702} (\bibinfo {year}
  {2010})}\BibitemShut {NoStop}%
\bibitem [{lyu()}]{lyutikov_sironi_komissarov_porth_2017}%
  \BibitemOpen
  \href@noop {} {\ \textbf {\bibinfo {volume} {83}}}\BibitemShut {NoStop}%
\bibitem [{\citenamefont {Finn}\ and\ \citenamefont
  {Kaw}(1977)}]{finn1977coalescence}%
  \BibitemOpen
  \bibfield  {author} {\bibinfo {author} {\bibfnamefont {J.~M.}\ \bibnamefont
  {Finn}}\ and\ \bibinfo {author} {\bibfnamefont {P.}~\bibnamefont {Kaw}},\
  }\href@noop {} {\bibfield  {journal} {\bibinfo  {journal} {The Physics of
  Fluids}\ }\textbf {\bibinfo {volume} {20}},\ \bibinfo {pages} {72} (\bibinfo
  {year} {1977})}\BibitemShut {NoStop}%
\bibitem [{\citenamefont {{Sweet}}(1958)}]{sweet_neutral_1958}%
  \BibitemOpen
  \bibfield  {author} {\bibinfo {author} {\bibfnamefont {P.~A.}\ \bibnamefont
  {{Sweet}}},\ }in\ \href@noop {} {\emph {\bibinfo {booktitle} {Electromagnetic
  Phenomena in Cosmical Physics}}},\ \bibinfo {series} {IAU Symposium},
  Vol.~\bibinfo {volume} {6},\ \bibinfo {editor} {edited by\ \bibinfo {editor}
  {\bibfnamefont {B.}~\bibnamefont {{Lehnert}}}}\ (\bibinfo {year} {1958})\ p.\
  \bibinfo {pages} {123}\BibitemShut {NoStop}%
\bibitem [{\citenamefont {{Parker}}(1957)}]{parker_sweet_1957}%
  \BibitemOpen
  \bibfield  {author} {\bibinfo {author} {\bibfnamefont {E.~N.}\ \bibnamefont
  {{Parker}}},\ }\href@noop {} {\bibfield  {journal} {\bibinfo  {journal}
  {Journal of Geophysical Research}\ }\textbf {\bibinfo {volume} {62}},\
  \bibinfo {pages} {509} (\bibinfo {year} {1957})}\BibitemShut {NoStop}%
\bibitem [{\citenamefont {{Loureiro}}\ \emph {et~al.}(2007)\citenamefont
  {{Loureiro}}, \citenamefont {{Schekochihin}},\ and\ \citenamefont
  {{Cowley}}}]{loureiro2007instability}%
  \BibitemOpen
  \bibfield  {author} {\bibinfo {author} {\bibfnamefont {N.~F.}\ \bibnamefont
  {{Loureiro}}}, \bibinfo {author} {\bibfnamefont {A.~A.}\ \bibnamefont
  {{Schekochihin}}}, \ and\ \bibinfo {author} {\bibfnamefont {S.~C.}\
  \bibnamefont {{Cowley}}},\ }\href@noop {} {\bibfield  {journal} {\bibinfo
  {journal} {Physics of Plasmas}\ }\textbf {\bibinfo {volume} {14}},\ \bibinfo
  {pages} {100703} (\bibinfo {year} {2007})}\BibitemShut {NoStop}%
\bibitem [{\citenamefont {Lapenta}(2008)}]{lapenta2008self}%
  \BibitemOpen
  \bibfield  {author} {\bibinfo {author} {\bibfnamefont {G.}~\bibnamefont
  {Lapenta}},\ }\href@noop {} {\bibfield  {journal} {\bibinfo  {journal}
  {Physical Review Letters}\ }\textbf {\bibinfo {volume} {100}},\ \bibinfo
  {pages} {235001} (\bibinfo {year} {2008})}\BibitemShut {NoStop}%
\bibitem [{\citenamefont {{Samtaney}}\ \emph {et~al.}(2009)\citenamefont
  {{Samtaney}}, \citenamefont {{Loureiro}}, \citenamefont {{Uzdensky}},
  \citenamefont {{Schekochihin}},\ and\ \citenamefont
  {{Cowley}}}]{samtaney2009formation}%
  \BibitemOpen
  \bibfield  {author} {\bibinfo {author} {\bibfnamefont {R.}~\bibnamefont
  {{Samtaney}}}, \bibinfo {author} {\bibfnamefont {N.~F.}\ \bibnamefont
  {{Loureiro}}}, \bibinfo {author} {\bibfnamefont {D.~A.}\ \bibnamefont
  {{Uzdensky}}}, \bibinfo {author} {\bibfnamefont {A.~A.}\ \bibnamefont
  {{Schekochihin}}}, \ and\ \bibinfo {author} {\bibfnamefont {S.~C.}\
  \bibnamefont {{Cowley}}},\ }\href@noop {} {\bibfield  {journal} {\bibinfo
  {journal} {Physical Review Letters}\ }\textbf {\bibinfo {volume} {103}},\
  \bibinfo {eid} {105004} (\bibinfo {year} {2009})}\BibitemShut {NoStop}%
\bibitem [{\citenamefont {{Bhattacharjee}}\ \emph {et~al.}(2009)\citenamefont
  {{Bhattacharjee}}, \citenamefont {{Huang}}, \citenamefont {{Yang}},\ and\
  \citenamefont {{Rogers}}}]{bhattacharjee2009fast}%
  \BibitemOpen
  \bibfield  {author} {\bibinfo {author} {\bibfnamefont {A.}~\bibnamefont
  {{Bhattacharjee}}}, \bibinfo {author} {\bibfnamefont {Y.-M.}\ \bibnamefont
  {{Huang}}}, \bibinfo {author} {\bibfnamefont {H.}~\bibnamefont {{Yang}}}, \
  and\ \bibinfo {author} {\bibfnamefont {B.}~\bibnamefont {{Rogers}}},\
  }\href@noop {} {\bibfield  {journal} {\bibinfo  {journal} {Physics of
  Plasmas}\ }\textbf {\bibinfo {volume} {16}},\ \bibinfo {eid} {112102}
  (\bibinfo {year} {2009})}\BibitemShut {NoStop}%
\bibitem [{\citenamefont {Huang}\ and\ \citenamefont
  {Bhattacharjee}(2010)}]{huang2010scaling}%
  \BibitemOpen
  \bibfield  {author} {\bibinfo {author} {\bibfnamefont {Y.-M.}\ \bibnamefont
  {Huang}}\ and\ \bibinfo {author} {\bibfnamefont {A.}~\bibnamefont
  {Bhattacharjee}},\ }\href@noop {} {\bibfield  {journal} {\bibinfo  {journal}
  {Physics of Plasmas}\ }\textbf {\bibinfo {volume} {17}},\ \bibinfo {pages}
  {062104} (\bibinfo {year} {2010})}\BibitemShut {NoStop}%
\bibitem [{\citenamefont {{Uzdensky}}\ \emph {et~al.}(2010)\citenamefont
  {{Uzdensky}}, \citenamefont {{Loureiro}},\ and\ \citenamefont
  {{Schekochihin}}}]{uzdensky2010fast}%
  \BibitemOpen
  \bibfield  {author} {\bibinfo {author} {\bibfnamefont {D.~A.}\ \bibnamefont
  {{Uzdensky}}}, \bibinfo {author} {\bibfnamefont {N.~F.}\ \bibnamefont
  {{Loureiro}}}, \ and\ \bibinfo {author} {\bibfnamefont {A.~A.}\ \bibnamefont
  {{Schekochihin}}},\ }\href@noop {} {\bibfield  {journal} {\bibinfo  {journal}
  {Physical Review Letters}\ }\textbf {\bibinfo {volume} {105}},\ \bibinfo
  {pages} {235002} (\bibinfo {year} {2010})}\BibitemShut {NoStop}%
\bibitem [{\citenamefont {{Loureiro}}\ \emph {et~al.}(2012)\citenamefont
  {{Loureiro}}, \citenamefont {{Samtaney}}, \citenamefont {{Schekochihin}},\
  and\ \citenamefont {{Uzdensky}}}]{loureiro2012magnetic}%
  \BibitemOpen
  \bibfield  {author} {\bibinfo {author} {\bibfnamefont {N.~F.}\ \bibnamefont
  {{Loureiro}}}, \bibinfo {author} {\bibfnamefont {R.}~\bibnamefont
  {{Samtaney}}}, \bibinfo {author} {\bibfnamefont {A.~A.}\ \bibnamefont
  {{Schekochihin}}}, \ and\ \bibinfo {author} {\bibfnamefont {D.~A.}\
  \bibnamefont {{Uzdensky}}},\ }\href@noop {} {\bibfield  {journal} {\bibinfo
  {journal} {Physics of Plasmas}\ }\textbf {\bibinfo {volume} {19}},\ \bibinfo
  {pages} {042303} (\bibinfo {year} {2012})}\BibitemShut {NoStop}%
\bibitem [{\citenamefont {{Loureiro}}\ \emph
  {et~al.}(2013{\natexlab{a}})\citenamefont {{Loureiro}}, \citenamefont
  {{Schekochihin}},\ and\ \citenamefont {{Zocco}}}]{loureiro2013fast}%
  \BibitemOpen
  \bibfield  {author} {\bibinfo {author} {\bibfnamefont {N.~F.}\ \bibnamefont
  {{Loureiro}}}, \bibinfo {author} {\bibfnamefont {A.~A.}\ \bibnamefont
  {{Schekochihin}}}, \ and\ \bibinfo {author} {\bibfnamefont {A.}~\bibnamefont
  {{Zocco}}},\ }\href@noop {} {\bibfield  {journal} {\bibinfo  {journal}
  {Physical Review Letters}\ }\textbf {\bibinfo {volume} {111}},\ \bibinfo
  {pages} {025002} (\bibinfo {year} {2013}{\natexlab{a}})}\BibitemShut
  {NoStop}%
\bibitem [{\citenamefont {{Loureiro}}\ and\ \citenamefont
  {{Uzdensky}}(2016)}]{loureiro_magnetic_2016}%
  \BibitemOpen
  \bibfield  {author} {\bibinfo {author} {\bibfnamefont {N.~F.}\ \bibnamefont
  {{Loureiro}}}\ and\ \bibinfo {author} {\bibfnamefont {D.~A.}\ \bibnamefont
  {{Uzdensky}}},\ }\href@noop {} {\bibfield  {journal} {\bibinfo  {journal}
  {Plasma Physics and Controlled Fusion}\ }\textbf {\bibinfo {volume} {58}},\
  \bibinfo {eid} {014021} (\bibinfo {year} {2016})}\BibitemShut {NoStop}%
\bibitem [{Note1()}]{Note1}%
  \BibitemOpen
  \bibinfo {note} {In the collisionless case $\beta _{\protect \rm rec}\simeq
  0.1$~\cite {cassak2017review} should also remain constant in time, even
  though the reconnection regime transitions from laminar to plasmoid-mediated
  as $R_n$ grows while the ion skin depth $d_i$ remains constant~\cite
  {ji2011phase} --- though the efficiency of coalescence may be decreased~\cite
  {karimabadi2011flux,stanier2015role}.}\BibitemShut {Stop}%
\bibitem [{\citenamefont {Biskamp}\ and\ \citenamefont
  {Welter}(1989)}]{biskamp1989dynamics}%
  \BibitemOpen
  \bibfield  {author} {\bibinfo {author} {\bibfnamefont {D.}~\bibnamefont
  {Biskamp}}\ and\ \bibinfo {author} {\bibfnamefont {H.}~\bibnamefont
  {Welter}},\ }\href@noop {} {\bibfield  {journal} {\bibinfo  {journal}
  {Physics of Fluids B: Plasma Physics}\ }\textbf {\bibinfo {volume} {1}},\
  \bibinfo {pages} {1964} (\bibinfo {year} {1989})}\BibitemShut {NoStop}%
\bibitem [{\citenamefont {Olesen}()}]{olesen2015inverse}%
  \BibitemOpen
  \bibfield  {author} {\bibinfo {author} {\bibfnamefont {P.}~\bibnamefont
  {Olesen}},\ }\href@noop {} {\bibinfo  {journal} {arXiv preprint
  arXiv:1509.08962}\ }\BibitemShut {NoStop}%
\bibitem [{\citenamefont {Burgers}(1948)}]{burgers1948mathematical}%
  \BibitemOpen
\bibfield  {journal} {  }\bibfield  {author} {\bibinfo {author} {\bibfnamefont
  {J.~M.}\ \bibnamefont {Burgers}},\ }in\ \href@noop {} {\emph {\bibinfo
  {booktitle} {Advances in applied mechanics}}},\ Vol.~\bibinfo {volume} {1}\
  (\bibinfo  {publisher} {Elsevier},\ \bibinfo {year} {1948})\ pp.\ \bibinfo
  {pages} {171--199}\BibitemShut {NoStop}%
\bibitem [{\citenamefont {Kadomtsev}\ and\ \citenamefont
  {Pogutse}()}]{kadomtsev1973nonlinear}%
  \BibitemOpen
  \bibfield  {author} {\bibinfo {author} {\bibfnamefont {B.}~\bibnamefont
  {Kadomtsev}}\ and\ \bibinfo {author} {\bibfnamefont {O.}~\bibnamefont
  {Pogutse}},\ }\href@noop {} {\bibfield  {journal} {\bibinfo  {journal} {Sov.
  Phys. JETP}\ }\textbf {\bibinfo {volume} {5}},\ \bibinfo {pages}
  {575}}\BibitemShut {NoStop}%
\bibitem [{\citenamefont {Strauss}(1976)}]{strauss1976nonlinear}%
  \BibitemOpen
  \bibfield  {author} {\bibinfo {author} {\bibfnamefont {H.~R.}\ \bibnamefont
  {Strauss}},\ }\href@noop {} {\bibfield  {journal} {\bibinfo  {journal} {The
  Physics of Fluids}\ }\textbf {\bibinfo {volume} {19}},\ \bibinfo {pages}
  {134} (\bibinfo {year} {1976})}\BibitemShut {NoStop}%
\bibitem [{\citenamefont {{Schekochihin}}\ \emph {et~al.}(2009)\citenamefont
  {{Schekochihin}}, \citenamefont {{Cowley}}, \citenamefont {{Dorland}},
  \citenamefont {{Hammett}}, \citenamefont {{Howes}}, \citenamefont
  {{Quataert}},\ and\ \citenamefont
  {{Tatsuno}}}]{schekochihin2009astrophysical}%
  \BibitemOpen
  \bibfield  {author} {\bibinfo {author} {\bibfnamefont {A.~A.}\ \bibnamefont
  {{Schekochihin}}}, \bibinfo {author} {\bibfnamefont {S.~C.}\ \bibnamefont
  {{Cowley}}}, \bibinfo {author} {\bibfnamefont {W.}~\bibnamefont {{Dorland}}},
  \bibinfo {author} {\bibfnamefont {G.~W.}\ \bibnamefont {{Hammett}}}, \bibinfo
  {author} {\bibfnamefont {G.~G.}\ \bibnamefont {{Howes}}}, \bibinfo {author}
  {\bibfnamefont {E.}~\bibnamefont {{Quataert}}}, \ and\ \bibinfo {author}
  {\bibfnamefont {T.}~\bibnamefont {{Tatsuno}}},\ }\href@noop {} {\bibfield
  {journal} {\bibinfo  {journal} {The Astrophysical Journal Supplement Series}\
  }\textbf {\bibinfo {volume} {182}},\ \bibinfo {pages} {310} (\bibinfo {year}
  {2009})}\BibitemShut {NoStop}%
\bibitem [{\citenamefont {Loureiro}\ \emph {et~al.}(2016)\citenamefont
  {Loureiro}, \citenamefont {Dorland}, \citenamefont {Fazendeiro},
  \citenamefont {Kanekar}, \citenamefont {Mallet}, \citenamefont {Vilelas},\
  and\ \citenamefont {Zocco}}]{loureiro2016viriato}%
  \BibitemOpen
  \bibfield  {author} {\bibinfo {author} {\bibfnamefont {N.~F.}\ \bibnamefont
  {Loureiro}}, \bibinfo {author} {\bibfnamefont {W.}~\bibnamefont {Dorland}},
  \bibinfo {author} {\bibfnamefont {L.}~\bibnamefont {Fazendeiro}}, \bibinfo
  {author} {\bibfnamefont {A.}~\bibnamefont {Kanekar}}, \bibinfo {author}
  {\bibfnamefont {A.}~\bibnamefont {Mallet}}, \bibinfo {author} {\bibfnamefont
  {M.}~\bibnamefont {Vilelas}}, \ and\ \bibinfo {author} {\bibfnamefont
  {A.}~\bibnamefont {Zocco}},\ }\href@noop {} {\bibfield  {journal} {\bibinfo
  {journal} {Computer Physics Communications}\ }\textbf {\bibinfo {volume}
  {206}},\ \bibinfo {pages} {45} (\bibinfo {year} {2016})}\BibitemShut
  {NoStop}%
\bibitem [{Note2()}]{Note2}%
  \BibitemOpen
  \bibinfo {note} {The number of islands is numerically determined by
  diagnosing the O-points and X-points of the system, identified with the
  maximum/minimum and saddle points of $\psi (x,y)$~\cite
  [e.g.,][]{servidio2009magnetic}.}\BibitemShut {Stop}%
\bibitem [{\citenamefont {{Loureiro}}\ \emph
  {et~al.}(2013{\natexlab{b}})\citenamefont {{Loureiro}}, \citenamefont
  {{Schekochihin}},\ and\ \citenamefont {{Uzdensky}}}]{loureiro2013plasmoid}%
  \BibitemOpen
  \bibfield  {author} {\bibinfo {author} {\bibfnamefont {N.~F.}\ \bibnamefont
  {{Loureiro}}}, \bibinfo {author} {\bibfnamefont {A.~A.}\ \bibnamefont
  {{Schekochihin}}}, \ and\ \bibinfo {author} {\bibfnamefont {D.~A.}\
  \bibnamefont {{Uzdensky}}},\ }\href@noop {} {\bibfield  {journal} {\bibinfo
  {journal} {Physical Review E}\ }\textbf {\bibinfo {volume} {87}},\ \bibinfo
  {pages} {013102} (\bibinfo {year} {2013}{\natexlab{b}})}\BibitemShut
  {NoStop}%
\bibitem [{Note3()}]{Note3}%
  \BibitemOpen
  \bibinfo {note} {We also observe in our simulations that kinetic energy
  decays as $u^2\sim t^{-1}$, consistent with $u\sim B$.}\BibitemShut {Stop}%
\bibitem [{\citenamefont {Cassak}\ \emph {et~al.}(2017)\citenamefont {Cassak},
  \citenamefont {Liu},\ and\ \citenamefont {Shay}}]{cassak2017review}%
  \BibitemOpen
  \bibfield  {author} {\bibinfo {author} {\bibfnamefont {P.}~\bibnamefont
  {Cassak}}, \bibinfo {author} {\bibfnamefont {Y.-H.}\ \bibnamefont {Liu}}, \
  and\ \bibinfo {author} {\bibfnamefont {M.}~\bibnamefont {Shay}},\ }\href@noop
  {} {\bibfield  {journal} {\bibinfo  {journal} {Journal of Plasma Physics}\
  }\textbf {\bibinfo {volume} {83}} (\bibinfo {year} {2017})}\BibitemShut
  {NoStop}%
\bibitem [{\citenamefont {Ji}\ and\ \citenamefont
  {Daughton}(2011)}]{ji2011phase}%
  \BibitemOpen
  \bibfield  {author} {\bibinfo {author} {\bibfnamefont {H.}~\bibnamefont
  {Ji}}\ and\ \bibinfo {author} {\bibfnamefont {W.}~\bibnamefont {Daughton}},\
  }\href@noop {} {\bibfield  {journal} {\bibinfo  {journal} {Physics of
  Plasmas}\ }\textbf {\bibinfo {volume} {18}},\ \bibinfo {pages} {111207}
  (\bibinfo {year} {2011})}\BibitemShut {NoStop}%
\bibitem [{\citenamefont {{Karimabadi}}\ \emph {et~al.}(2011)\citenamefont
  {{Karimabadi}}, \citenamefont {{Dorelli}}, \citenamefont {{Roytershteyn}},
  \citenamefont {{Daughton}},\ and\ \citenamefont
  {{Chac{\'o}n}}}]{karimabadi2011flux}%
  \BibitemOpen
  \bibfield  {author} {\bibinfo {author} {\bibfnamefont {H.}~\bibnamefont
  {{Karimabadi}}}, \bibinfo {author} {\bibfnamefont {J.}~\bibnamefont
  {{Dorelli}}}, \bibinfo {author} {\bibfnamefont {V.}~\bibnamefont
  {{Roytershteyn}}}, \bibinfo {author} {\bibfnamefont {W.}~\bibnamefont
  {{Daughton}}}, \ and\ \bibinfo {author} {\bibfnamefont {L.}~\bibnamefont
  {{Chac{\'o}n}}},\ }\href@noop {} {\bibfield  {journal} {\bibinfo  {journal}
  {Physical Review Letters}\ }\textbf {\bibinfo {volume} {107}},\ \bibinfo
  {eid} {025002} (\bibinfo {year} {2011})}\BibitemShut {NoStop}%
\bibitem [{\citenamefont {Stanier}\ \emph {et~al.}(2015)\citenamefont
  {Stanier}, \citenamefont {Daughton}, \citenamefont {Chacon}, \citenamefont
  {Karimabadi}, \citenamefont {Ng}, \citenamefont {Huang}, \citenamefont
  {Hakim},\ and\ \citenamefont {Bhattacharjee}}]{stanier2015role}%
  \BibitemOpen
  \bibfield  {author} {\bibinfo {author} {\bibfnamefont {A.}~\bibnamefont
  {Stanier}}, \bibinfo {author} {\bibfnamefont {W.}~\bibnamefont {Daughton}},
  \bibinfo {author} {\bibfnamefont {L.}~\bibnamefont {Chacon}}, \bibinfo
  {author} {\bibfnamefont {H.}~\bibnamefont {Karimabadi}}, \bibinfo {author}
  {\bibfnamefont {J.}~\bibnamefont {Ng}}, \bibinfo {author} {\bibfnamefont
  {Y.-M.}\ \bibnamefont {Huang}}, \bibinfo {author} {\bibfnamefont
  {A.}~\bibnamefont {Hakim}}, \ and\ \bibinfo {author} {\bibfnamefont
  {A.}~\bibnamefont {Bhattacharjee}},\ }\href@noop {} {\bibfield  {journal}
  {\bibinfo  {journal} {Physical Review Letters}\ }\textbf {\bibinfo {volume}
  {115}},\ \bibinfo {pages} {175004} (\bibinfo {year} {2015})}\BibitemShut
  {NoStop}%
\bibitem [{\citenamefont {{Servidio}}\ \emph {et~al.}(2009)\citenamefont
  {{Servidio}}, \citenamefont {{Matthaeus}}, \citenamefont {{Shay}},
  \citenamefont {{Cassak}},\ and\ \citenamefont
  {{Dmitruk}}}]{servidio2009magnetic}%
  \BibitemOpen
  \bibfield  {author} {\bibinfo {author} {\bibfnamefont {S.}~\bibnamefont
  {{Servidio}}}, \bibinfo {author} {\bibfnamefont {W.~H.}\ \bibnamefont
  {{Matthaeus}}}, \bibinfo {author} {\bibfnamefont {M.~A.}\ \bibnamefont
  {{Shay}}}, \bibinfo {author} {\bibfnamefont {P.~A.}\ \bibnamefont
  {{Cassak}}}, \ and\ \bibinfo {author} {\bibfnamefont {P.}~\bibnamefont
  {{Dmitruk}}},\ }\href@noop {} {\bibfield  {journal} {\bibinfo  {journal}
  {Physical Review Letters}\ }\textbf {\bibinfo {volume} {102}},\ \bibinfo
  {eid} {115003} (\bibinfo {year} {2009})}\BibitemShut {NoStop}%
\end{thebibliography}%

\end{document}